\documentclass{aa} 
\usepackage[authoryear]{natbib}
\bibpunct{(}{)}{;}{a}{}{,}
\usepackage{amsmath}
\usepackage{graphicx}
\usepackage{lscape}

\begin{document}

\titlerunning{Testing the IAD$_0$-SPH concept} 
\authorrunning{Cabez\'on, Garc\'{i}a-Senz \& Escart\'in}

\title{Testing the concept of integral approach to derivatives within the smoothed particle hydrodynamics technique in astrophysical scenarios}

\author{Rub\'en M. Cabez\'on 
        \inst{1}, 
        Domingo Garc\'{i}a-Senz\inst{2} 
        \and
        Jos\'e A. Escart\'{i}n\inst{2}
       }
\institute
{Departement Physik. Universit\"at Basel. Klingelbergstrasse 82, 4056 Basel, Switzerland \\
 \and
Dept. de F\'{i}sica i Enginyeria Nuclear, Universitat Polit\`ecnica de Catalunya. Compte d'Urgell 187, 08036 Barcelona (Spain) and Institut d'Estudis Espacials de Catalunya. Gran Capit\` a 2-4, 08034 Barcelona, Spain \\}

\date{}

\abstract
{The smoothed particle hydrodynamics (SPH) technique is a well-known numerical method that has been applied to simulating the evolution of a wide variety of systems. Modern astrophysical applications of the method rely on the Lagrangian formulation of fluid Euler equations, which is fully conservative. A different scheme, based on a matrix approach to the SPH equations is currently being used in computational fluid dynamics (CDF). An original matrix formulation of SPH based on an integral approach to the derivatives, called IAD$_0$, has been recently proposed and is fully conservative and well-suited to simulating astrophysical processes.}  
{The behavior of the IAD$_0$ scheme is analyzed in connection with several astrophysical scenarios, and compared to the same simulations carried out with the standard SPH technique.} 
{The proposed hydrodynamic scheme is validated using a variety of numerical tests that cover important topics in astrophysics, such as the evolution of supernova remnants, the stability of self-gravitating bodies, and the coalescence of compact objects.}    
{The analysis of the hydrodynamical simulations of the above-mentioned astrophysical scenarios suggests that the SPH scheme built with the integral approach to the derivatives improves the results of the standard SPH technique. In particular, there is a better development of hydrodynamic instabilities, a good description of self-gravitating structures in equilibrium and a reasonable description of the process of coalescence of two white dwarfs. We also observed good conservations of energy and both linear and angular momenta that were generally better than those of standard SPH. In addition the new scheme is less susceptible to pairing instability.}
{We present a formalism based on a tensor approach to Euler SPH equations that we checked using a variety of three-dimensional tests of astrophysical interest. This new scheme is more accurate because of the re-normalization imposed on the interpolations, which is fully conservative and less prone to undergoing the pairing instability. The analysis of these test cases suggests that the method may improve the simulation of many astrophysical problems with only a moderate computational overload.}   


\maketitle

\section{Introduction}

Multidimensional numerical hydrodynamics is one of the most powerful tools of modern astrophysics to comprehend the cosmos machinery. Among them, the smoothed particle hydrodynamics (SPH) is one of the most widely used techniques because of its ability to describe the evolution of fluids with complicated geometries and a diversity of length scales. Since it was formulated, more than thirty years ago, by \cite{gm77} and \cite{lucy77}, it has largely evolved incorporating, little by little, a plethora of methods that makes it competitive compared to grid-based methods of Eulerian type. Details of the modern mathematical formulation, as well as of the main features of the state-of-art of the SPH technique, can be found in the reviews by \cite{monaghan05}, \cite{rosswog09},  \cite{springel10}, and \cite{price12}. 

\cite{gs12} (henceforth paper I) recently suggested that the use of matrix methods, \cite{dilts99}, in astrophysics could improve the simulations with the SPH technique with an affordable computational cost. In this work, we focus on specific three-dimensional (3D) astrophysical applications of the scheme formulated in paper I. In particular, we choose examples from different fields of astronomy to check the method and highlight its potential advantages over the standard SPH scheme. The suitability of IAD$_0$ for describing hydrodynamic instabilities found in paper I is confirmed by simulating the evolution of a supernova remnant (SNR). As the SNR evolves embedded in an uniform background of particles with negligible gravity, there are no numerical troubles affecting the outer limits of the system. The existence of boundaries becomes relevant to the second test, which is devoted to describing the equilibrium features of polytropes with different indexes and masses. For this problem, the interplay between pressure forces and gravity becomes crucial to ensure that the obtained structures are compatible with the analytical models. We show that the tensor method leads to polytropes where the central density and radius are slightly closer to the theoretical predictions than those obtained with the standard SPH technique. We also explore the ability of IAD$_0$ to describe a very dynamical situation by simulating the coalescence of two white dwarfs. In this case, a catastrophic merging of the stars ensues after a few orbital periods. For this test, the tensor method gives results of, at least, comparable quality to those obtained using the standard SPH scheme, but displaying a more homogeneous mixing of the material of both stars. There are also some differences in the angular velocity distribution of the remnant. For a similar elapsed time, the matrix calculation does not lead to the complete rigid rotation of the core, which is, however, achieved in the simulation using the standard scheme.  

The text is organized as follows. In Sec.~\ref{section2}, we review the mathematical formalism linked to IAD$_0$ and discuss its most relevant features. In Sec.~\ref{section3}, we describe the three astrophysical tests aimed at validating the code and comparing its performance to that of standard SPH. Section~\ref{section4} is devoted to incorporating thermal conductive transport in the tensor scheme, and to check the resulting algorithm. The benchmarking of the code is done in Sec.~\ref{section5}. Finally, the main conclusions of our work, as well as some comments about the shortcomings of the developed scheme and future lines of improvement, are outlined in the last section, which is devoted to our conclusions.

\section{Main features of the IAD$_0$ scheme}
\label{section2}

In paper I, it was shown that a conservative SPH scheme can be deduced from an integral approach to the derivatives. As starting point, we define the integral, 
  
\begin{equation}
I({\bf {r}})= \int_{V} \left[f({\bf {r'}})-f({\bf {r}})\right] ({\bf {r'}}-{\bf {r}}) W(\vert {\bf {r}'}-{\bf {r}}\vert, h) dr'{^3}\,,
\label{int}
\end{equation}

\noindent where $W(\vert {\bf {r}'}-{\bf {r}}\vert, h)$ is a spherically symmetric interpolating function and $h$ is called the smoothing length. The integral $I({\bf {r}})$ can be used to find the gradient of a function $f({\bf {r}})$ in a way similar to that used to evaluate the Laplace operator based on another integral expression in standard SPH. See \cite{brook85} and \cite{monaghan05}. The IAD$_0$ interpretation of SPH is the consequence of approaching Eq.~(\ref{int}) with summations along with two reasonable simplifications

\begin{equation}
f({\bf {r_b}})-f({\bf {r_a}})\simeq\bf\nabla f_a\cdot({\bf {r_b}}-{\bf {r_a}})\,,
\label{gradient}
\end{equation}

\noindent where $a$ and $b$ refer to neighboring particles with masses $m_a$ and $m_b$, respectively, and 

\begin{equation}
I({\bf {r_a}})\simeq
\sum_b\frac{m_b}{\rho_b} f({\bf r_b})({\bf {r_b}}-{\bf {r_a}}) W(\vert {\bf {r_b}}-{\bf {r_a}}\vert, h_a)
\label{approxI}
\end{equation}

\noindent is the corresponding discrete expression for Eq.~(\ref{int}). Note that, because the kernel is spherically symmetric, the factor $f({\bf {r_a}})$ does not appear in the expression above.

\noindent
The direct application of Eqs.~(\ref{int}),(\ref{gradient}), and (\ref{approxI}) to calculating the gradient of density, leads to a matrix equation

\begin{equation}
\left[
\begin{array}{c}
\partial \rho/\partial x_1\\
\partial \rho/\partial x_2\\
\partial \rho/\partial x_3\\
\end{array}
\right]_a
=
\left[
\begin{array}{ccc}
\tau_{11} & \tau_{12} & \tau_{13} \\
   \tau_{21}&\tau_{22}&\tau_{23} \\
   \tau_{31}&\tau_{32}&\tau_{33}
\end{array}
\right]^{-1}
\left[
\begin{array}{c}
I_1\\
I_2\\
I_3\\
\end{array}
\right]\,,
\label{matrix}
\end{equation}

\noindent where

\begin{equation}
\tau_{ij,a}=\sum_b \frac{m_b}{\rho_b}(x_{i,b}-x_{i,a})(x_{j,b}-x_{j,a})W_{ab}(h_a)\,; i,j=1,3
\label{tauijsph}
\end{equation}

\noindent and

\begin{equation}
I_{k,a}=\sum_b m_b~(x_{k,b}-x_{k,a}) W_{ab}(h_a)\,; k=1,3.
\end{equation}

It was shown in paper I that Eq.~(\ref{matrix}) leads to a formulation of the SPH Euler equations that is compatible with the variational principle. These equations are given by

\begin{equation}
\rho_a=\sum_{b=1}^{n_b} m_b W_{ab}(\vert {\bf r_b-r_a\vert},h_a)\,,
\label{density}
\end{equation}

\small
\begin{equation}
\ddot x_{i,a}=-\sum_{b=1}^{n_b} m_b\left(\frac{P_a}{\Omega_a\rho_a^2}\mathcal A_{i,ab}(h_a)+\frac{P_b}{\Omega_b\rho_b^2}\mathcal A'_{i,ab}(h_b)+\Pi_{ab}~\widetilde{\mathcal A}_{i,ab}\right)\,,
\label{momentumLqij}
\end{equation}

\begin{equation}
\left(\frac{du}{dt}\right)_a=\sum_{b=1}^{n_b}\sum_{i=1}^d m_b(v_{i,a}-v_{i,b})\left(\frac{P_a}{\Omega_a~\rho_a^2}~\mathcal{A}_{i,ab}(h_a)+ \frac{\Pi_{ab}}{2}~\widetilde{\mathcal A}_{i,ab}\right)\,,
\label{energyqij}
\end{equation}
\normalsize

\noindent where $\mathcal{A}_{i,ab}$ and $\mathcal{A}'_{i,ab}$ are 

\begin{align}
{\mathcal A}_{i,ab}(h_a) &=\sum_{j=1}^{d} c_{ij,a}(h_a) (x_{j,b}-x_{j,a}) W_{ab}(h_a)\,,\label{lagr7}
\end{align}
\begin{align}
{\mathcal A}'_{i,ab}(h_b) &=\sum_{j=1}^{d} c_{ij,b}(h_b) (x_{j,b}-x_{j,a}) W_{ab}(h_b)\,,\label{lagr8}
\end{align}

\noindent being $c_{ij}$ the coefficients of the inverse matrix defined in Eq.~(\ref{matrix}) and $d$ the dimension of the space. The magnitude $\Omega_{a}=(1-\partial\rho/\partial h~ \sum_a m_a~ \partial W_{ab}/\partial h)$ accounts for the gradient of the smoothing length. As usual, $\Pi_{ab}$ gives the viscous pressure due to the artificial viscosity (AV)

\begin{equation}
\Pi_{ab}=
\left\{\begin{array}{rclcc}
\frac{-\alpha c_{ab} \mu_{ab}+ \beta {\mu^2_{ab}}}{\bar\rho_{ab}} & \mathrm{for}~ & {\bf r}_{ab}\cdot {\bf v}_{ab} < 0, \\
 0~~ \mathrm{otherwise,}
\end{array}
\right.
\label{avis}
\end{equation}

\noindent and the remaining symbols have their usual meaning. The coefficient $\mu_{ab}$ is

\begin{equation}
\mu_{ab}=\frac{\bar h_{ab}{\bf r}_{ab}\cdot {\bf v}_{ab}}{r_{ab}^2+0.01~\bar h^2_{ab}}\,.
\end{equation}

\noindent To compute the viscous acceleration, the arithmetic mean of ${\mathcal A}$ is taken to be

\begin{equation}
\widetilde{\mathcal A}_{i,ab}=\frac{1}{2}\left[{\mathcal A}_{i,ab}(h_a)+{\mathcal A'}_{i,ab}(h_b)\right]\,.
\label{atilde}
\end{equation}

\noindent Therefore Eqs.~(\ref{density}),~(\ref{momentumLqij}) and (\ref{energyqij}) summarize the basis of the IAD$_0$ formalism. 

It is worth noting the strong similitudes between the above equations and those of standard SPH. The main difference is that vector expressions have been changed to tensor relationships that encode the renormalization of the many summations that appear in the SPH technique. Any expression of standard SPH can indeed be made compatible with IAD$_0$ by taking the kernel derivative as

\begin{equation}
\frac{\partial W_{ab}(h_a)}{\partial x_{i,a}}= {\mathcal A}_{i,ab}(h_a)\,; i=1,3\,,
\label{gradk}
\end{equation}   

\noindent where the coefficients ${\mathcal A}_{i,ab}(h_a)$ are defined by Eq.~(\ref{lagr7}). If the matrix coefficients in Eq.~(\ref{matrix}) are calculated analytically, the matrix $\mathcal T$ becomes diagonal. In this case, it can be shown that for Gaussian kernels the standard and IAD descriptions are totally equivalent. Hereafter, the scheme obtained from the diagonal form of $\mathcal T$ is referred to as {\sl vector-}IAD$_0$.

\subsection{Harmonic kernels}

\begin{figure}
\includegraphics[angle=-90,width=\columnwidth]{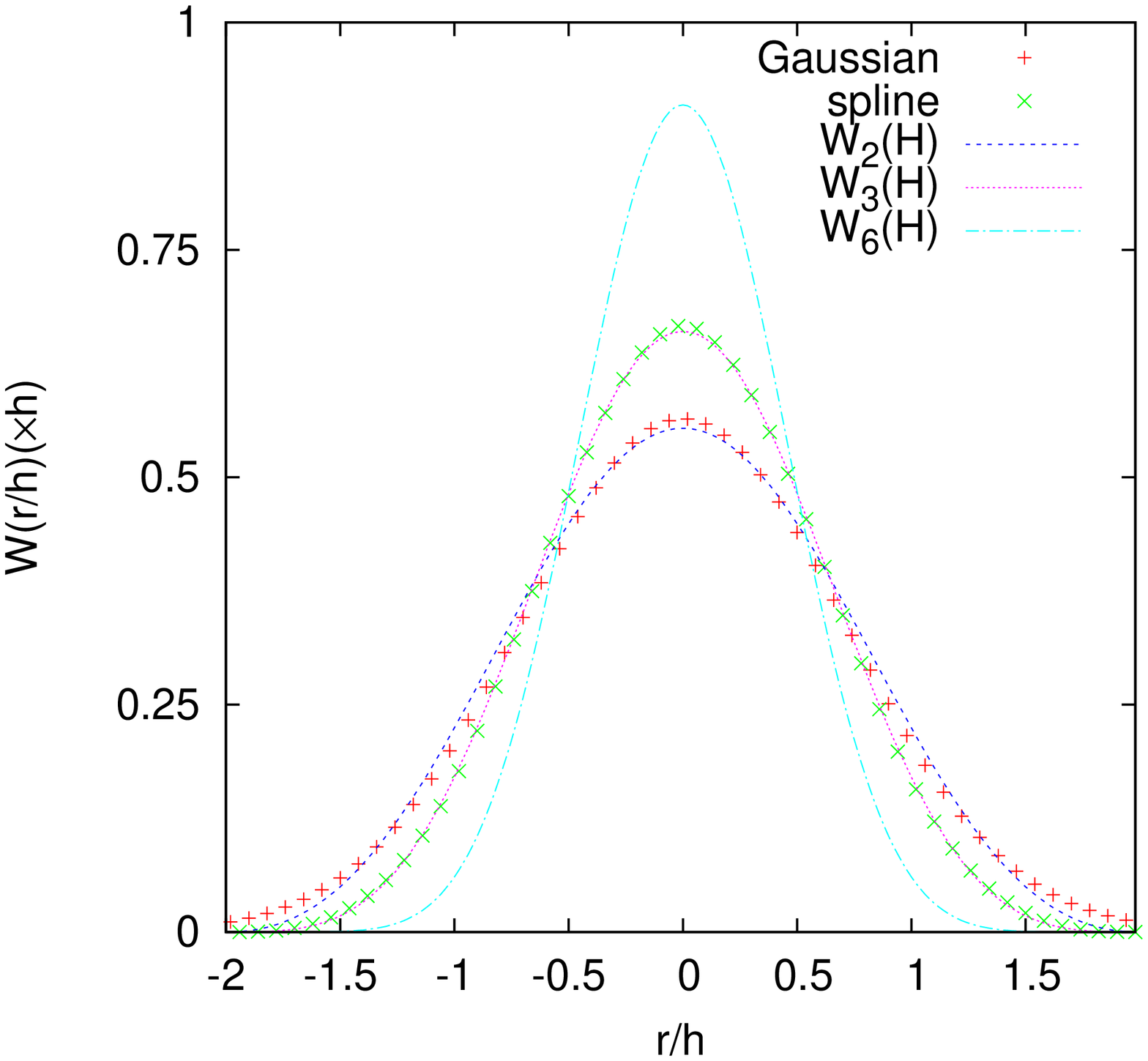}
\caption{Profile of several kernels in one dimension. The Gaussian kernel is truncated at $r=2h$. The $W_n(H)$ belongs to the harmonic family of kernels described by Eq.~(\ref{harmonic}) for $n=2$, 3, and 6, respectively.} 
\label{figure1}
\end{figure}

All the simulations presented in this work were carried out using the so-called harmonic kernels devised by \cite{cabezon08}, to perform the SPH interpolations. Being relatively unknown kernels with very interesting features, their use warrants some explanation. They are defined as

\begin{equation}
 W_n^H(v,h)= \frac{B_n}{h^3}~{\mathrm sinc^n\left(\frac{\pi}{2} v\right)}\,;\quad  0\leq v\leq 2\,,
\label{harmonic}
\end{equation}   

\noindent where $\mathrm sinc(u) = \frac{\sin u}{u}$, $n$ is the index of the kernel, and $B_n$ is the normalization constant. By defining $\mathrm sinc(0)=1$, the function $\mathrm sinc$ is extended to an analytical function with compact support on the real axis. Because of its connection with spectral analysis, the function $sinc(u)$ is of special relevance to signal analysis, from where it borrows its name. The profile of the kernel for several values of the leading index $n$ is shown in Fig.~\ref{figure1}, where it can be seen that the profile of $W_3^H(v,h)$ closely matches that of the cubic spline. The behavior of $W_3^H(v,h)$ is therefore very similar to that of the cubic spline, with the advantage that its second derivative is continuous and can be differentiated many times. Switching the index to $n=2$ gives the $W_2^H(v,h)$ kernel, with a profile close to that of the truncated Gaussian kernel, as shown in Fig.~\ref{figure1}. The implementation of the $W_n^H(v,h)$ family of kernels adds more flexibility to SPH, as one can, for example, take a different index $n$ to handle the artificial viscosity terms in either the momentum and energy equations or the heat conduction equation, without changing the number of neighbors of the particle. They are also useful for avoiding the pairing instability (see Sec.~\ref{pairing}) because whenever two particles approach each other too closely, the index of the kernel can be increased to block the development of the instability.

The normalization constants $B_n$ for a large number of values of $n$ were calculated in \cite{cabezon08}, where a fitting analytical formula for $B_n$ was also provided. To increase the computational speed, it is recommended that the value of ${\mathrm sinc\left(\frac{\pi}{2} v\right)}$ and its derivative be stored in a table as a function of \mbox{$v,\ (0\le v\le 2)$}, and that a linear Taylor expansion be used to calculate the value of $\mathrm sinc$. This allows a fast computation of Eq.~(\ref{harmonic}) and its derivative once the index $n$ has been chosen.

\subsection{Pairing instability}
\label{pairing}
The simulations oriented to benchmark the different SPH schemes reported in this manuscript as well as in paper I, suggest that there is an additional advantage of the matrix method. In general, calculations carried out using IAD$_0$ are less affected by the pairing instability. For a given interpolating kernel with spherical symmetry, there is a fiducial distance to the center $r_0$ at which the first derivative of the kernel reaches its maximum absolute value. Within this critical radius, a pair of neighboring particles feels an increasingly weaker repulsive force, which can lead to the artificial clumping of the particles (but see \cite{deh12} for a different explanation of the origin of this instability). The exact location of $r_0$ depends on the number of neighbors and the peculiarities of the kernel. For either the cubic-spline kernel or the harmonic kernel with index n=3, its location is $r_0=\frac{2}{3}h$, while for more centrally condensed kernels it is at yet smaller radii.

\begin{figure}
\includegraphics[angle=-90,width=\columnwidth]{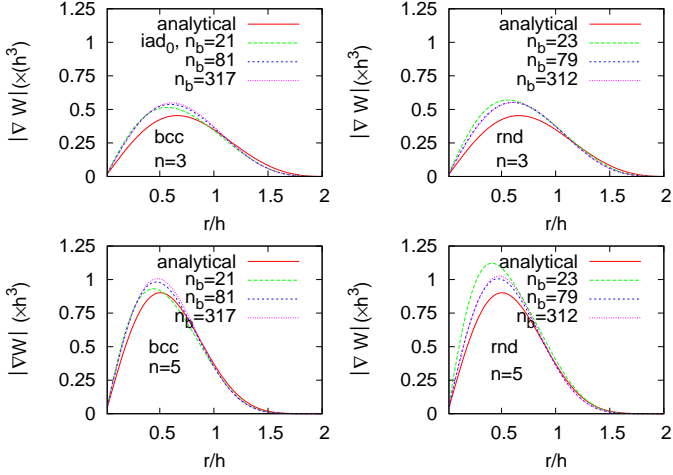}
\caption{Profile of the gradient of the kernel calculated analytically in the standard SPH way (referred as {\sl analytical} in the figures) as well as numerically, through Eq.~(\ref{gradk}) for two particle settings: {\sl bcc} and pseudo-random, kernel indexes $n=3$, 5, and different number of neighbors.} 
\label{figure2}
\end{figure}

The robustness of matrix methods to cope with pairing instability can be understood by confronting the analytical estimation of the gradient of the kernel within the standard framework, to the numerical value obtained using Eq.~(\ref{gradk}). The numerical derivative of $W_{ab}$ for the particle $a$ located at the center of a 2D lattice was evaluated using two different particle settings and harmonic kernel indexes. Figure~\ref{figure2} shows the value of $\nabla_a W_{ab}$ at different distances from the kernel origin. The left column stands for a {\sl bcc}-type particle setting in a two-dimensional square lattice, while the column on the right represents a quasi-uniform distribution of particles, obtained by randomly perturbing the {\sl bcc} distribution with a maximum perturbation amplitude of $0.3\Delta$, where $\Delta$ is the lattice spacing. The harmonic kernel index was set to $n=3$ (thus, reproducing  the cubic-spline kernel) and to $n=5$ to see the effect of making the profile sharper. It can be seen that increasing $n$ (keeping h constant) always raises the maximum of the kernel derivative, as expected for a more centrally condensed kernel, shifting the abscissa where the maximum is achieved closer to zero. For $n=3$ and $n=5$, the maximum is taken at $r/h=2/3$ and $r/h=0.504$, respectively. The effect of changing the numerical scheme also has an impact on the kernel derivative. According to Fig.~\ref{figure2}, the  maximum of the gradient of the kernel always appears closer to the origin than in the standard scheme, regardless of the number of neighbors. Moreover, the absolute value of the maximum is larger for IAD$_0$. Together, this means that the tensor scheme is less affected by pairing instability than the standard SPH scheme.
   
\subsection{Free surface conditions}
\label{freesurface}
Handling boundaries is often a difficult point of the SPH technique. In static gaseous configurations, such as stars or planets, the radius of the object is the result of the careful balance between gravity and pressure forces. The gradient in the pressure is, however, not as accurately estimated by SPH near the surface as in the star's interior. Therefore, the radius of the configuration is a magnitude not as well-defined as others once the mechanical equilibrium is achieved.

\begin{figure}
\includegraphics[angle=-90,width=\columnwidth]{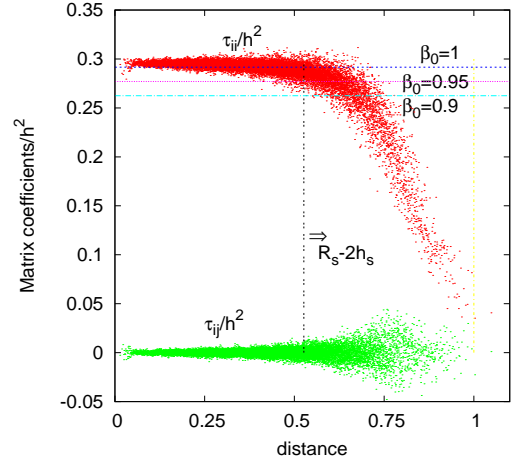}
\caption{Profile of coefficients $\tau_{ii}/h^2$ and $\tau_{ij}/h^2, (j\neq i)$ in a polytrope of index $n=5/2$ fitted with $20,000$ particles (model F$_1$ of Table~\ref{table2}). The distance to the center is normalized to the radius of the polytrope, $R_s$. The analytical value of $\tau_{ii}/h^2$ is also given (horizontal dashed line corresponding to $\beta=1$). Basically, all the star interior  has a value $\beta\ge 0.95$, while beyond a radius $R_s-2h_s$ (where $h_s$ is the smoothing length close to the surface) the value of $\beta$ rapidly declines (see Sec.~\ref{freesurface} for the definition of $\beta$).} 
\label{figure3}
\end{figure}

Unfortunately, the direct application of IAD$_0$ to free surface boundaries does not solve the problem. The reason is that the magnitudes $\tau_{ij}$ given by Eq.~(\ref{tauijsph}), which serve to normalize the derivatives, are very sensitive to the presence of boundaries. Close to the edge of the system, the gradient of pressure is overestimated and the equilibrium of the body is reached at a larger radius than in the standard (STD) scheme. The impact of the free boundaries on the $\tau_{ij}$ coefficients is well-illustrated in Fig.~\ref{figure3}, which depicts the profile of these magnitudes along the polytropic structure discussed in Sec.~\ref{polytropes}. The numerical estimation of $\tau_{ij}$ can be compared to their analytical value calculated using
  
\begin{equation}
\begin{split}
\tau_{ii}^a=&\frac{1}{3}\int_{\infty} r^2 W(v)dr^3=\\
&\frac{4\pi}{3}B_n h^2\int_0^2 v^4 sinc^n(\frac{\pi}{2}v)~dv\,;i=1,3\,,
\end{split}
\label{inttauii}
\end{equation} 
 
\noindent which for $n=3$ gives $\tau_{11}^a=\tau_{22}^a=\tau_{33}^a=0.2916~h^2$. We see that, with the exception of the surface layers, the analytical and the numerical values agree with an accuracy better than a 5\%. These coefficients are, however, very sensitive to boundaries and rapidly decay when the first particles belonging to the surface enter the summations. This led some authors \citep[e.g.][]{oger07} to propose a practical algorithm for choosing the adequate value of $\tau_{ii}$ as
      
\begin{equation}
\tau_{ii}=
\left\{\begin{array}{lcl}
\tau_{ii}    &(\mathrm{Eq}.~\ref{tauijsph})&  \qquad \beta \geq \beta_0\,,\\
\tau_{ii}^a  &                             &  \qquad \mathrm{otherwise,}
\end{array}
\right.
\label{tauijselect}
\end{equation}

\noindent where $\beta=\tau_{ii}/\tau_{ii}^a$ and $\beta_0\simeq 0.95$. In Sec.~\ref{polytropes}, we discuss the impact of the different approaches to $\tau_{ii}$ in describing the structure of polytropes with different indexes. We chose several values for $\beta_0$ ranging from $\beta_0=0$ (fully tensor IAD$_0$) to $\beta_0=\infty$ ({\sl vector-}IAD$_0$). The values $\beta_0=0.90$ and $\beta_0=0.97$ suggested by \cite{oger07}  were also checked. Our main conclusion is that the $\beta_0=0$ fully tensor IAD$_0$ scheme provides the best description of the structure of these polytropes, although the numerical noise is larger than for the STD scheme. Interestingly, the vector-like approach with $\beta_0=\infty$ give results as good as the STD scheme with the same computational cost. Nevertheless, an additional advantage of the full tensor scheme is that it improves the description of the growth of hydrodynamic instabilities, as suggested in the tests described in Sec.~\ref{section3} and paper I. On another note, we have not seen any particular advantage of using a hybrid scheme with $\beta_0\simeq 0.95$ to handle self-gravitating bodies, which has the negative side of increasing the numerical noise. Furthermore, a value of $\beta<1$ is not necessarily linked to a free boundary because it may appear, for example, in the presence of mildly or strong shock waves. For these reasons, we conclude that the renormalized IAD$_0$ scheme summarized by Eqs.~(\ref{tauijsph}) to (\ref{lagr8}) should be preferentially used to carry out simulations of astrophysical systems.

\section{Astrophysical tests}
\label{section3}

A basic check of the new scheme was described in paper I. For the most part, the technique was tested in two dimensions with particles located in ordered lattices and, sometimes, using particles with different masses. For these specific tests, the IAD$_0$ scheme showed a good behavior because it was able to handle the Rayleigh-Taylor and Kelvin-Helmholtz instabilities better than standard SPH. Simulation of supersonic events (Sedov blast wave and wall-heating shock) were of similar quality, if not slightly better, than those computed using STD smoothed particle hydrodynamics. The total energy was always more well-conserved when the tensor method was used.

The only 3D calculation in paper I was that used to simulate the evolution towards stability of a Sun-like polytrope. In that simulation, gravity was calculated using a multipolar expansion of the force (\cite{hern89}). The main conclusion of the test was that, although good equilibrium configurations were achieved in both methods, in the tensor method more energy is stored as numerical noise for the same elapsed time.

To check the code in realistic astrophysical scenarios, we chose three simulations related to different hydrodynamic processes. These include the growth of the RT instability in supernova remnants, the stability of polytropes of different masses and polytropic indexes, and the study of the coalescence process of a pair of compact stars. The calculations were carried out by choosing $n=3$ in the harmonic kernels given in Eq.~\ref{harmonic}. Unless explicitly stated, the number of neighbors was kept constant to $n_b=100$ in all simulations. Precise details of the initial particle setting and physics included are presented in each subsection of the corresponding test.

\subsection{Hydrodynamics of a supernova remnant} 
\label{remnant}
The evolution of supernova remnants has to be studied in more than one dimension to capture the fine details of their structure. Several things may contribute to cause the evolution to deviate from the spherical symmetry: the lack of symmetry of the exploding object that gives rise to the SNR, the inhomogeneities in the ambient medium (AM), or the interaction of the remnant with cosmic rays \cite[]{wang11}. One of the physical phenomenae that was among the first to be studied as a source of spatial irregularities in SNRs was the Rayleigh-Taylor (RT) instability. In SNR, the RT instability may often develop in the region between the forward and reverse shocks, induced by the deceleration of the supernova shell. Pioneering 3D simulations of the impact of the RT instability on the evolution of the remnant were conducted by \cite{che92}, by taking a small slice of the full domain to enhance the resolution. More recently, 3D simulations encompassing the whole SNR were developed by \cite{vigh11}. To reproduce the growth and structure of the instability, a large amount of computational cells are needed. Two-dimensional simulations carried out by \cite{dwa00} suggest that a minimum of 300x300 mesh zones (in 2D) have to be used to describe the growth of the RT fingers. Therefore, the ability to simulate the growth of the RT instability in SNR is a good test for 3D hydrocodes.  

All multi-dimensional simulations performed so far have been carried out using grid-based codes, generally of Eulerian type. In spite of its wide use in astrophysics, SPH has scarcely been applied to the study of SNR. One exception was the study of \cite{gsbs12}, who used an axisymmetric SPH code to study the imprint of the secondary star on the geometry of the remnant resulting from a Type Ia supernova explosion. It is worth noting that SPH is able to describe the growth of the RT instabilities in 3D, using a smaller number of particles than those suggested in \cite{dwa00} ($N\simeq 1.5~10^6$ particles instead of $2-3~10^7$ grid cells) because there is no waste of computational resources in the lowest density regions of the simulated domain.    

To perform the SNR simulation, we built an exponential profile for the ejecta following the prescriptions given in \cite{dwa00}. The profile was mapped to a 3D set of equal-mass particles and allowed to interact with an homogeneous ambient medium. A perfect-gas equation of state (EOS), where $\gamma=5/3$ was assumed. 

The density of the ejecta was

\begin{equation}
\rho_{ej}(t)= A \exp\left(-\frac{v}{v_e}\right)~t^{-3}\,,
\label{exp}
\end{equation}

\noindent where the constants $A$ and $v_e$ are

\begin{equation}
A= 7.67~10^6~\mathrm {g~cm^{-3}~s^3}\left(\frac{M_{ej}}{M_{Ch}}\right)^\frac{5}{2} E_{51}^{-\frac{3}{2}}\,,
\end{equation}     

\begin{equation}
v_e=2.44~10^8~\mathrm{cm~s^{-1}}~E_{51}^{\frac{1}{2}}\left(\frac{M_{ej}}{M_{Ch}}\right)^{-\frac{1}{2}}\,.
\end{equation}
             
We took $M_{ej}=M_{Ch}$ and $E_{51}=1$ in the expressions above. The time $t$ in Eq.~(\ref{exp}) was set to $t_0=2\cdot 10^9$ s, and the velocity profile at $t_0$ was assumed to be homologous, with $v(r)=r/t_0$. With these choices, the initial size of the ejecta was $\simeq 1$ pc and the density at the outer edge matched that of the AM medium (assumed here to be homogeneous and have a value $\rho_{AM}=2\cdot 10^{-24}$ g.cm$^{-3}$). The 1D density profile of the ejecta was mapped onto a 3D sample of particles spread across a sphere of radius 1 pc, according to the radial density profile given by Eq.~(\ref{exp}), and where the angular location of the particles was chosen at random. This random setting led to a blurred density profile, which was finally driven to the spherical symmetry by means of a tangential relaxation of the model using SPH. To do this, we allowed the particles to move under the action of pressure forces, although their movement was constrained to ensure that the distance to the center remained constant. The homogeneous AM was simply reproduced by sampling the particles in a cubic-centered, {\sl bcc} grid with a size of 10~pc. Using $N_{ej}=257,776$ particles and $N_{AM}=1,206,576$ particles, the mass ratio of particles belonging to the AM to those in the ejecta is $m_{ej}/m_{AM}\simeq 2.3$. To check that such a mass contrast has a negligible impact on the results, we recalculated model B$_1$ of Table~\ref{table1} using a number of particles for the ambient medium $N_{AM}'=2.3 N_{AM}$. There were no significant changes in the evolution of the model.

The evolution of the SNR was simulated for different initial conditions using both the STD and the IAD$_0$ codes, with the goal of analyzing their performance. 

A single mode perturbation in the initial radial-velocity field was seeded at $t=0$~yr according to the expression

\begin{equation}
\Delta v_r = \delta~(1-\exp(-(r/R_{ej})^2))~\cos(n~\theta)\,,
\label{vpert}
\end{equation}

\noindent where $R_{ej}=1$ pc, $\theta$ is the azimuthal angle, $\theta=\cos^{-1}(z/r)$, and the parameter $\delta$ is close to the maximum perturbation velocity we wish to impose. The wavenumber was set to $n=12$ in all models. 

\begin{figure}
\includegraphics[angle=0,width=\columnwidth]{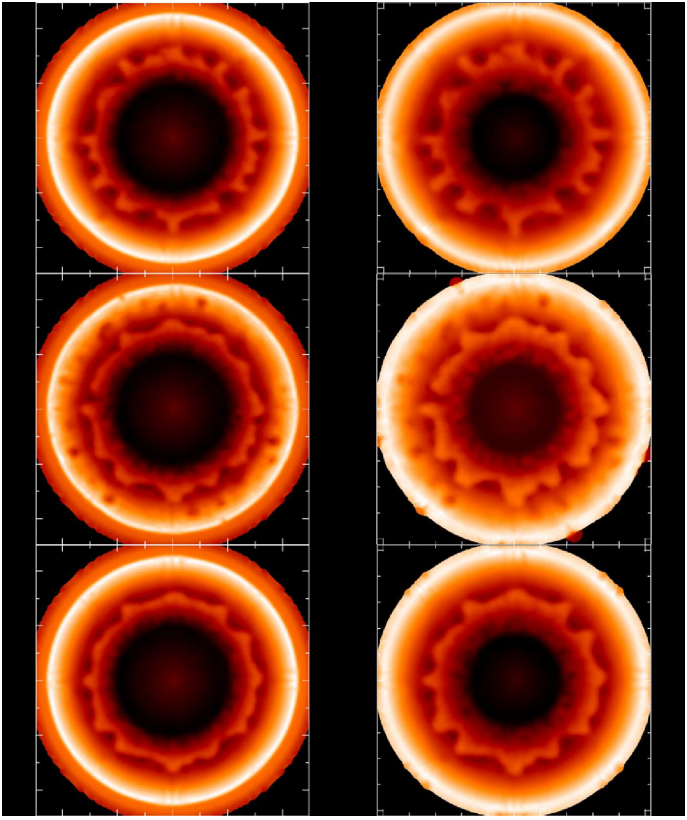}
\caption{Density color map depicting the growth of the RT instability in a SNR for models B$_1$ (upper panels), B$_2$ (middle panels), and B$_3$ (lower panels) of Table~\ref{table1} at times $t=698$~yr (left columns) and $t=951$~yr (right columns). The size of the box in all panels is 10 pc in each direction.} 
\label{figure4}
\end{figure}

The inspection of the results summarized in Table~\ref{table1} reveals that energy is always better conserved, by a factor $\simeq 2$, when the tensor method is used. Linear and angular momentum conservations are at least as good as in the standard formulation, independently of either the absence (models A$_1$ and A$_2$) or the presence (models B$_1$ and B$_2$) of the velocity perturbation. Figure~\ref{figure4} presents a color map of density in the YZ plane at $t=698$~yr for models B$_1$ and B$_2$ of Table~\ref{table1}, calculated by assuming $\delta=500$~km.s$^{-1}$ in Eq.~(\ref{vpert}). The combined effect of the radial velocity perturbation and the anisotropies of the $bcc$ lattice leads to the growth of the RT instability. At $t=698$~yr, the development of the instability is already appreciably larger in the tensor calculation than in the STD scheme. The differences between both calculations become more pronounced at $t=951$~yr, when the forward shock reaches the limits of the system. These results agree with the main conclusions of paper I, in terms of the development of the RT instability in 2D stratified fluids inside a homogeneous gravitational field.

\begin{figure}
\includegraphics[angle=-90,width=\columnwidth]{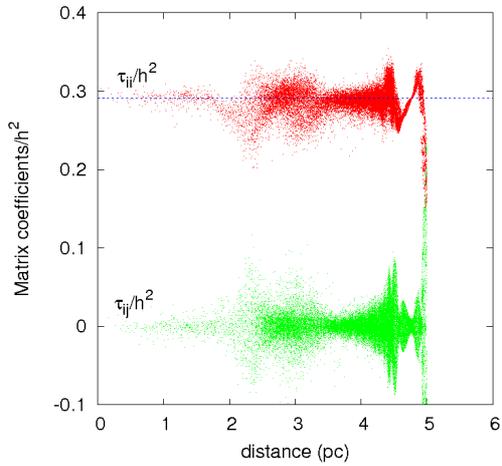}
\caption{Profile of the coefficients $\tau_{ii}/h^2$ and $\tau_{ij}/h^2, (j\neq i)$ for model B$_1$ of Table~\ref{table1}. The analytical value of $\tau_{ii}/h^2$ is also given (horizontal dashed line).} 
\label{figure5}
\end{figure}

Model B$_3$ in Table~\ref{table1} and the lower panels of Fig.~\ref{figure4} refer to {\sl vector-}IAD$_0$. As we can see, the evolution of the RT instability is similar to that of the STD scheme but total energy is better conserved and the density map seems to be slightly cleaner. Therefore, we conclude that the renormalization of the derivatives, which characterizes the fully tensor IAD$_0$ method, makes it more suitable for describing the evolution of hydrodynamic instabilities. The importance of renormalization is highlighted in Fig.~\ref{figure5}, which depicts the profile of the elements $\tau_{ii}$ and $\tau_{ij}$ calculated with Eq.~(\ref{tauijsph}). As we can see, there is a large deviation of $\tau_{ii}$ from its analytical estimation through Eq.~(\ref{inttauii}). Similarly, the off-diagonal matrix elements are not equal to zero. The re-normalization of the derivative then helps the growth of the instability, confirming the main conclusions given in paper I.

\begin{figure}
\includegraphics[angle=-90,width=\columnwidth]{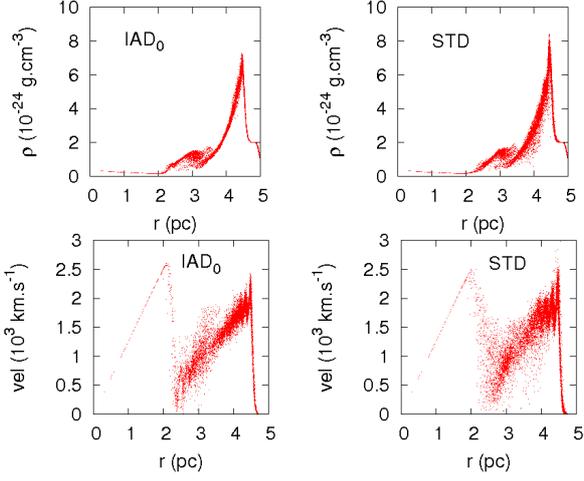}
\caption{Density and velocity profiles of the SNR at elapsed time $t=698$~yr for the stretched-grid models of the supernova ejecta (models C$_1$ and C$_2$ in Table~\ref{table1}). Outside the RT-unstable region, the spherical symmetry is better preserved when the tensor method is used.} 
\label{figure6}
\end{figure}

As a final test, we built an developed model for the supernova ejecta using the stretched-grid method (\cite{he92}, \cite{gs98}). These models are labeled C$_1$ and C$_2$ in Table~\ref{table1}. This kind of initial models displays almost perfect spherically symmetric density profiles, but the price to pay is a large deformation of the particle lattice. Although no perturbation was seeded, the larger anisotropies of the grid soon lead to the growth of the RT instability in the region between the forward and reverse shocks. The results of the simulations are shown in Fig.~\ref{figure6}. The upper panels depict the density profile after $t=698$~yr  of evolution, whereas the bottom ones show the radial velocity profile at the same elapsed time. We see that outside the region where the RT instability develops the profiles calculated using IAD$_0$ show less dispersion than those computed with the STD scheme. In the RT unstable layer located between the forward and reverse shocks, the dispersion is larger, as expected. In addition, there is a factor of two enhancement in the energy conservation when the tensor method is used (last two rows in Table~\ref{table1}).

\subsection{Polytropes of index n=3/2 and n=5/2 in equilibrium}
\label{polytropes}
Gaseous self-gravitating structures can usually be roughly approached using a polytrope with the appropriate index. Because of their simplicity and known theoretical properties, polytropes are often used to benchmark multidimensional hydrocodes \cite[]{ste93,pri07}. We explored the abilities of the tensor scheme in reproducing the equilibrium structure of polytropes of index $n=3/2$ and $n=5/2$, and confronted the results with the predictions of standard SPH models. A mass of $M=0.6 M_{\sun}$ and a radius of $R=8\cdot 10^8$~cm were assigned to the $n=3/2$ polytrope, so that it could represent a stable white dwarf with central density $\rho_c=3.3\cdot 10^6$~g.cm$^{-3}$. For the $n=5/2$ polytrope, we assigned it a mass of $M=1.35 M_{\sun}$ and a radius of $2\cdot 10^8$~cm, which are both representative of a massive white dwarf with central density $\rho_c=1.9\cdot 10^9$~g.cm$^{-3}$. Note that in this last case the mass is close to the Chandrasekhar-mass limit for a polytrope with $n=3$. This second structure is then much more unstable than the $0.6 M_{\sun}$ star. The initial models were built by randomly spreading $2\cdot 10^4$ particles in 3D, according to the mass density profile of the polytrope. In a second step, the models were relaxed in the non-radial direction to enhance their spherical symmetry. Afterwards, the particles were allowed to evolve freely in the 3D domain. The EOS was that of a perfect gas $P=\rho(\gamma-1)u$, with $\gamma=5/3$, $7/5$ for $n=3/2$, $5/2$ polytropes, respectively. The main features of both polytropes are summarized in Table~\ref{table2}. 

\begin{figure}
\includegraphics[angle=-90,width=\columnwidth]{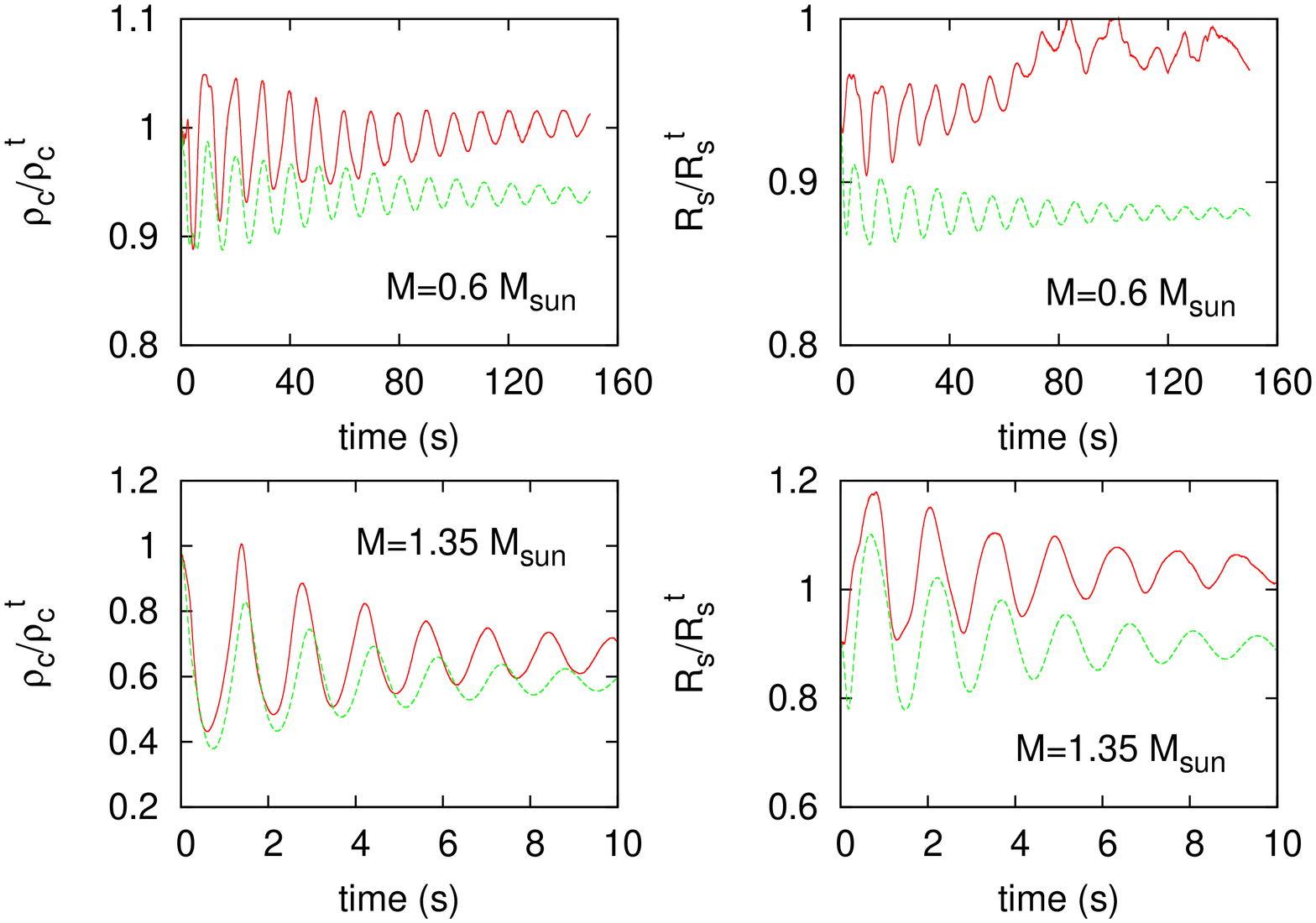}
\caption{Evolution of central density and surface radius (normalized to the theoretical values) for polytropes with indexes $n=3/2$ (upper panels, corresponding to models E$_1$ and E$_2$ in Table~\ref{table2}) and $n=5/2$ (lower panels, corresponding to models F$_1$ and F$_2$ in Table~\ref{table2}). Continuum and dashed lines are for the {\sl tensor-}IAD$_0$ and standard schemes, respectively.}
\label{figure7}
\end{figure}

Figure~\ref{figure7} represents the evolution of the central density and stellar radius for models E and F of Table~\ref{table2}. To smooth fluctuations, the central density and the radius were averaged over the closest particles to the center and the surface, respectively. The central density of the less massive structure converges to the analytical value after ten oscillatory cycles. The radius of the polytrope follows a similar evolution, although the value obtained with STD is $\simeq 90\%$ of the theoretical estimation owing to the small number of particles used in the simulation. We see that despite the lower convergence rate of the tensor calculation the final value of $\rho_c$ and $R$ are closer to the correct value than for the STD calculation. The relative error in the central density is $\epsilon_r(\rho_c)< 2\%$ for the IAD$_0$ calculation and a little higher, $\epsilon_r(\rho_c)\simeq 5\%$ for STD. For the more massive polytrope, the differences between the SPH schemes become more accentuated, as shown in the bottom panels of Fig.~\ref{figure7}. The fluctuation in the central density is much larger than in the previous case and the convergence towards the analytical value is less rapid. Nevertheless, the tensor calculation provides a more accurate estimation of the central density of the equilibrium configuration. According to Fig.~\ref{figure7}, the relative errors in the central density at the last calculated models are $\epsilon_r (\rho_c)\simeq  33\%$ and $\epsilon_r (\rho_c)\simeq  42\%$ for the tensor and standard methods, respectively. It is worth noting that a lower particle clumping was observed when the matrix scheme was used, which may be the origin of the small differences in the value of the central density calculated with both methods.

\begin{figure}
\includegraphics[angle=-90,width=\columnwidth]{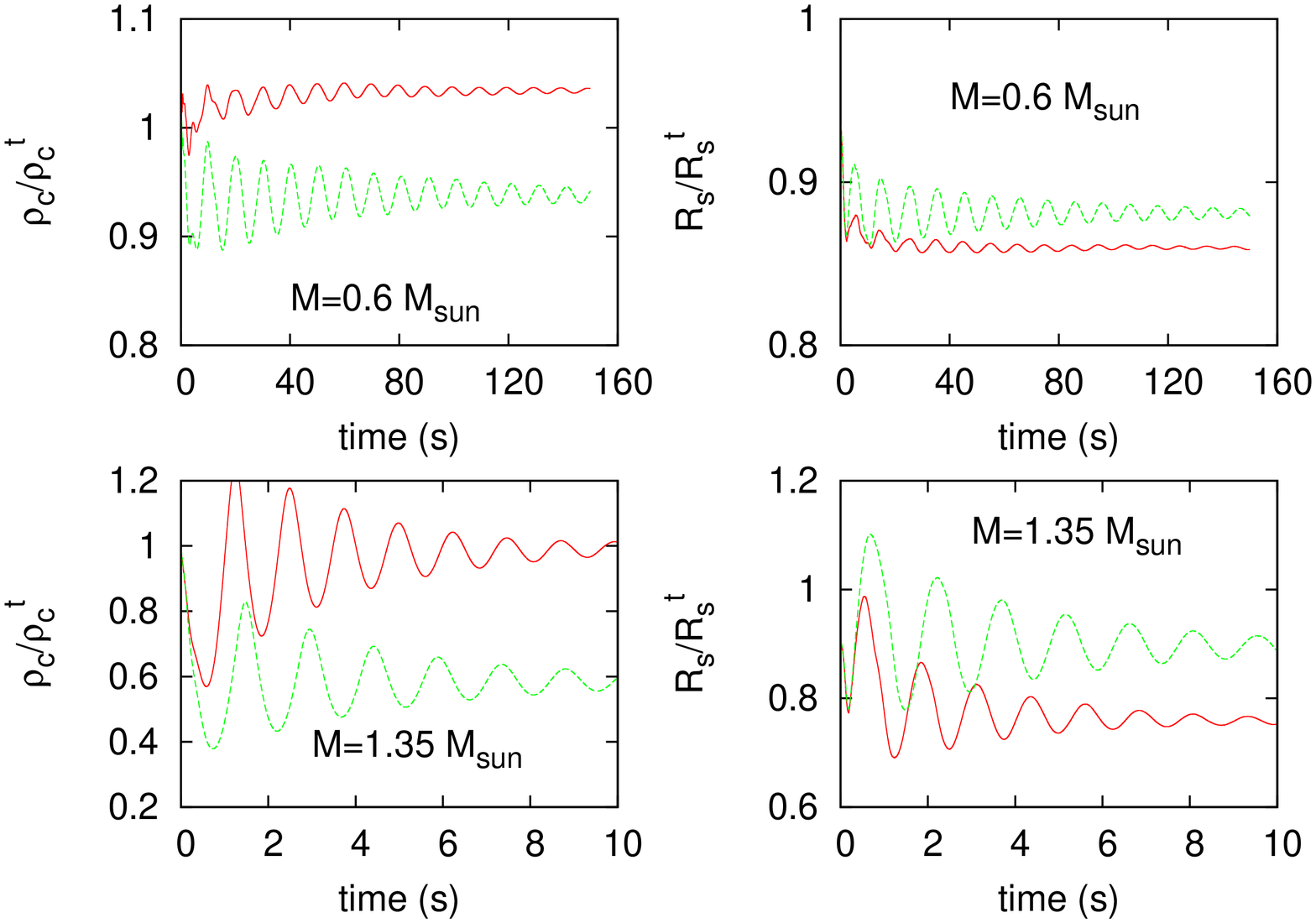}
\caption{Evolution of central density and surface radius (normalized to the theoretical values) for polytropes with indexes $n=3/2$ (upper panels, corresponding to models E$_2$ and E$_3$ in Table~\ref{table2}) and $n=5/2$ (lower panels, corresponding to models F$_2$ and F$_3$). Continuous and dashed lines are for {\sl vector-}IAD$_0$ and standard schemes, respectively.}
\label{figure8}
\end{figure}

Figure~\ref{figure8} shows the evolution of both polytropes calculated with {\sl vector-}IAD$_0$. The central density is much more closely reproduced than for the STD scheme but the radius of the configurations differs considerably from the actual value. Therefore, we conclude that the best combination of central density and surface radius is achieved by the full tensor IAD$_0$ scheme.

\begin{figure}
\includegraphics[angle=-90,width=\columnwidth]{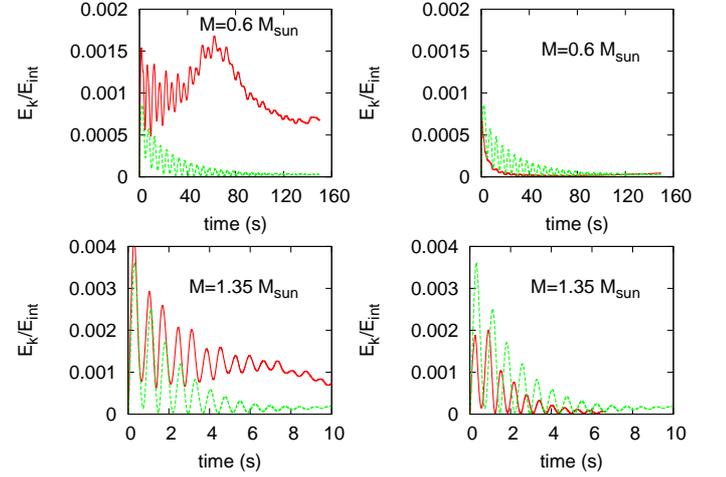}
\caption{Evolution of the ratio of kinetic to internal energies. Left column is for polytropes with indexes $n=3/2$ (upper panel) and $n=5/2$ (lower panel) calculated using {\sl tensor-}IAD$_0$ (continuum line) and STD (dashed line). Same for pictures on the right column, where, however, {\sl vector-}IAD$_0$ was used.}
\label{figure9}
\end{figure}

The leftmost panels of Fig.~\ref{figure9} depict the evolution of the ratio of the kinetic to internal energies, $E_k/E_{int}$, of both polytropes. The evolution of the kinetic energy is that of a damped system when standard SPH is used, while the IAD$_0$ calculation is less dissipative. Therefore, the amount of kinetic energy stored in the configuration after several evolution cycles remains larger for the matrix calculation. All this points to the renormalization of the derivatives as the main agent behind the increase in numerical noise. To clarify this point, we ran two simulations, models E$_3$ and F$_3$ in Table~\ref{table2}, using the vector approach to IAD$_0$. The evolution of the kinetic energy is shown in the rightmost panels of Fig.~\ref{figure9}. As we can see, the damping of the systems is even faster that in the STD scheme and the final kinetic energy is lower. 

The level of energy conservation at the last calculated model is correlated with the amount of kinetic energy. According to the last column of Table~\ref{table2}, energy is most accurately conserved for the vector approach to IAD$_0$. For the $n=3/2$ polytrope, there is a better conservation for STD than for tensor IAD$_0$ but for the most unstable $n=5/2$ polytrope the level of conservation is similar.

\subsection{Merging of two white dwarfs}
\label{merger}

Because of its mesh-free features, a substantial amount of applications of SPH have been devoted to the study of the coalescence of stellar objects. In this section we simulated the merging of two twin polytropes of index n=3/2 and mass $0.6 M_{\sun}$ with both the IAD$_0$ and STD schemes. In each calculation, the structure was that of the last calculated model corresponding to models E$_1$ and E$_2$ in Table~\ref{table2}. The EOS used was that of a perfect gas with $\gamma=5/3$. Both stars were put in a circular rigid-rotation orbit in the plane XY with radius $r_{orb}^0=1.5~R_{s}$, where $R_{s}=8,000$~km is the theoretical surface radius of the polytrope, from the center of mass of the system. To enforce the coalescence, a braking force proportional to the velocity ${\bf F}_{bra}=-k(t)\cdot {\bf v}$ was imposed during one revolution period $P$, according to the prescription

\begin{equation}
 k(t)= 
\left\{\begin{array}{rclcc}
  \frac{0.1}{(P+t)} \, \qquad ~& t \leq P\,, \\
 $ 0 $ \,\qquad ~& t > P\,,
\end{array}
\right.
\label{braking}
\end{equation}

\noindent where $t$ is the elapsed time and  $P=29.3$~s. 

This scenario roughly accounts for the merging of twin white dwarfs described by an EOS dominated by the electrons in the partially degenerated regime.
   
\begin{figure}
\includegraphics[width=\columnwidth]{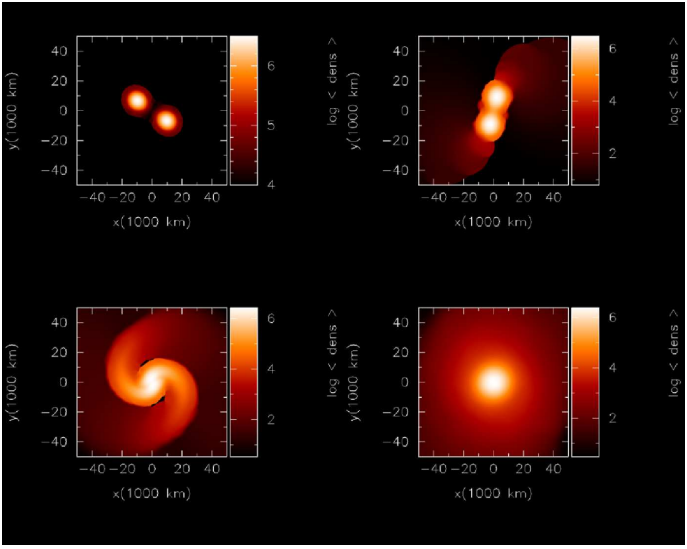}
\caption{Density color map of the coalescence process of two twin polytropes of index $n=3/2$ at times $t=0.31 P$, $t=2.8 P$, $t=4.3 P$, and $t=7.7 P$, ($P=29.3$ s) calculated using IAD$_0$.}
\label{figure10}
\end{figure}

\begin{figure}
\includegraphics[width=\columnwidth]{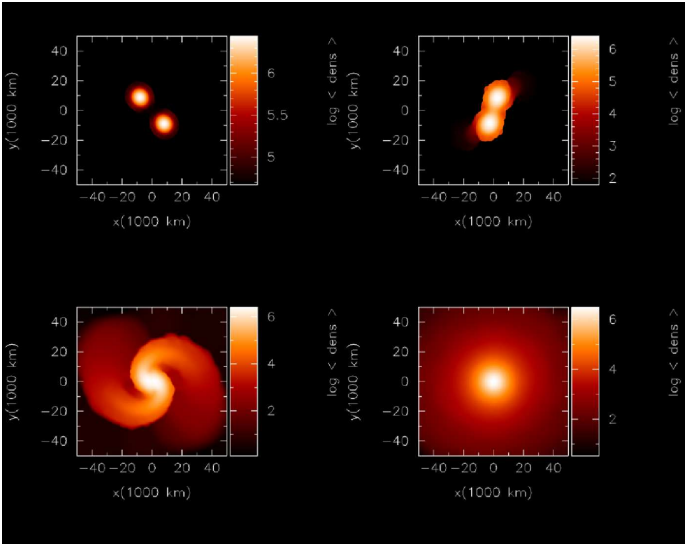}
\caption{Density color map of the coalescence process of two twin polytropes of index $n=3/2$ at times $t=0.23 P$, \mbox{$t=2.15 P$}, $t=3.3 P$, and $t=7.4 P$ ($P=29.3$ s) calculated using the standard SPH scheme.}

\label{figure11}
\end{figure}

\begin{figure}
\includegraphics[angle=-90,width=\columnwidth]{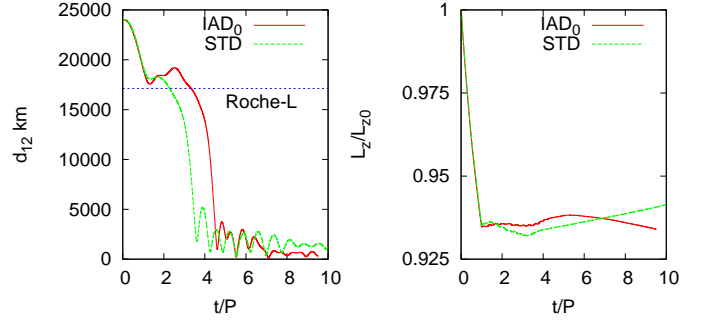}
\caption{Left: Evolution of the distance between the center of mass of both polytropes. Right: Evolution of the orbital angular momentum, $L_z$, normalized to its initial value. An artificial braking force was acting before $t/P=1$}
\label{figure12}
\end{figure}

\begin{figure}
\includegraphics[width=\columnwidth]{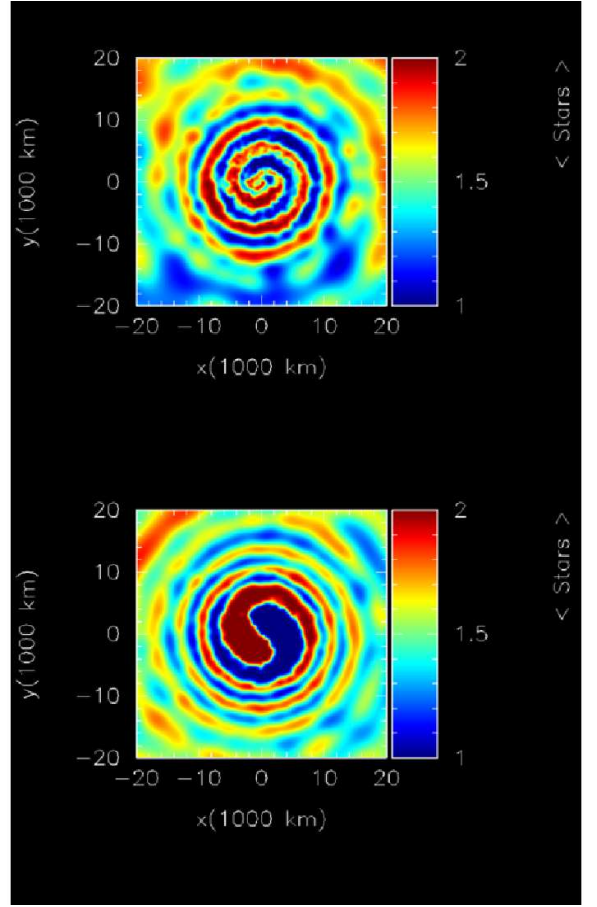}
\caption{Mixing of material of both stars in the core as calculated by IAD$_0$ at $t=227$~s (upper picture) and with STD at $t=216$~s (bottom picture). Blue and red colors refer to the gas belonging to each component of the original binary system. Mixed regions display a superposition of both colors.}
\label{figure13}
\end{figure}

\begin{figure}
\includegraphics[angle=-90,width=\columnwidth]{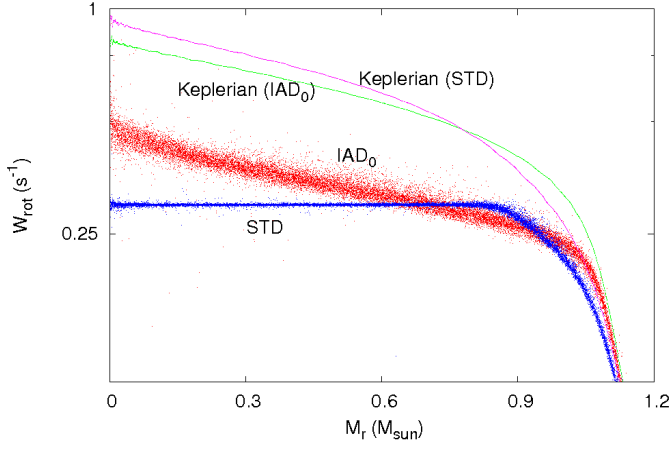}
\caption{Angular velocity profiles at $t\simeq 220$~s as function of mass coordinate, calculated using IAD$_0$ (red) and STD (blue), respectively. The Keplerian values for both calculations are given for reference.}
\label{figure14}
\end{figure}

The results of the simulations are summarized in Figs.~\ref{figure10} to \ref{figure14}. Figures~\ref{figure10} and \ref{figure11} represent a density color map of both stars in the orbital plane as obtained with IAD$_0$ and STD schemes, respectively. On the whole, the behavior is rather similar, although the coalescence evolves more slowly for the matrix method. We have to keep in mind, however, that the details of the evolution depend on the precise initial conditions \cite[]{dan11}. The initial equilibrium models for both calculations are similar, but not identical, because of the differences introduced during the relaxation process. The model E$_1$ of Table~\ref{table2}, approached with the tensor method, has a larger radius and less clumping than model E$_2$ calculated using the STD scheme. The details of the coalescence are also influenced by the errors in the estimation of gravity introduced by the tree-walk approach. These errors induce small differences that grow during the most dynamic phase of the merging. Thus, it is not straightforward to discern which part of the differences comes from the particular SPH scheme used in the simulation. On the other hand, both methods give the correct radius for Roche-Lobe (RL) overflow once the center of mass separation, $d_{12}$, becomes similar to $d_{12}=17,100$~km. Once the RL is crossed, the mass transfer becomes catastrophic and the merging of both stars takes place in a small fraction of one period.

Figure~\ref{figure12} shows the evolution of the $L_z$ component of the angular momentum, orthogonal to the orbital plane. For $t<P$, there is a constant loss of angular momentum due to the imposed external braking force. At times $t>P$, but before the dynamical phase of the merging ($t/P\simeq 4$), the tensor scheme displays a rather good conservation of $L_z$, with small ripples induced by the multipolar approach to gravity. Conservation of $L_z$ in this phase is not so good for the standard formulation. On another note, during the most violent phase of the coalescence and further bouncing of the core, the evolution of $L_z$ follows a slightly different trend in both schemes. The tensor method leads to a monotonic increase in $L_z$ until a maximum is reached followed by a steady decline. In the STD calculation, the behavior is the opposite. Nevertheless, the evolution  of the angular momentum after the merging is complicated by the formation of an extended halo of particles, which is clearly seen in the last snapshot in Figs.~\ref{figure10} and \ref{figure11}. Although only a tiny amount of mass belongs to the halo, such mass produces a large torque on the core and any error in the estimation of gravity translates into a larger error in the angular momentum.  

An important challenge for the hydrocodes is to adequately represent the mix of the advected material during the coalescence process. In our system, the initial conditions are fully symmetric. Therefore, one would expect that a few minutes after the merging the core is homogeneously composed of the material of both stars. Figure~\ref{figure13} depicts the approximate distribution of the gas belonging to each star one minute after the catastrophic stage of the coalescence process. As we can see, the material of both white dwarfs is much more thoroughly mixed in the IAD$_0$ calculation than the standard one. The merged core is made of successive thin, onion-like layers. In the calculation with the standard scheme, the onion-like structure is less pronounced, especially in the inner 10~km. We stress that the recipe to handle the artificial viscosity is exactly the same in both calculations. The only difference is the treatment of the gradient of the kernel, which seems ultimately to be the responsible for the larger amount of mixing obtained in the IAD$_0$ calculation.           

Figure~\ref{figure14} shows the angular velocity profile of the particles in the orbital plane as a function of their mass coordinate. As we can see, the standard calculation leads to an almost perfect rigid rotation for $M_r<0.8 M_{\sun}$, followed by a Keplerian velocity distribution beyond that point. This behavior agrees with the results obtained by \cite{ras12} for a similar scenario. At distances $M_r>1 M_{\sun}$, the velocity profile obtained with IAD$_0$ is also Keplerian, but below that mass, rigid rotation is never attained in the matrix calculation. In the standard calculation, the high amount of viscosity, which prevents the mixing of the core, now couples the different layers of the fluid so that the system rapidly approaches rigid rotation. Nevertheless, the time delay for core synchronization in nature is a function of the real physical viscosity. In the hydrodynamic models, the synchronization time is probably artificially shortened by the much larger numerical viscosity introduced by the codes. The impact of artificial viscosity in the final structure of the merged object was extensively discussed in \cite{dan11}.

\section{Adding thermal conduction  to IAD$_0$}
\label{section4}

It is not difficult to incorporate physics into the proposed matrix formulation of SPH by simply using Eq.~(\ref{gradk}) to calculate the gradient of the kernel whenever necessary. As an example, we work out an expression that can be used to calculate the conductive heat transport and apply it to simulate the evolution of a thermal wave. To do this, it is enough to consider the SPH thermal conduction equation \citep{springel10}

\begin{equation}
\frac{du_a}{dt}=\sum_{b=1}^{n_b} \frac{m_b}{\rho_a~\rho_b}\frac{(\kappa_a+\kappa_b)(T_b-T_a)}{r_{ab}^2} {\bf r}_{ab}\cdot {\bf\nabla}_a\tilde W_{ab}\,,
\label{conducstd}
\end{equation}    

\noindent where $\kappa$ is the thermal conductivity. If we estimate the gradient of the kernel using Eq.~(\ref{gradk}), it leads to,

\begin{equation}
\frac{du_a}{dt}=\sum_{b=1}^{n_b}\frac{m_b}{\rho_a~\rho_b}\frac{(\kappa_a+\kappa_b)(T_b-T_a)}{r_{ab}^2}\sum_{i=1}^d (x_{i,a}-x_{i,b})~\widetilde{\mathcal A}_{i,ab}\,, 
\label{conduciad}
\end{equation}  

\noindent where $d$ is the dimension of the space and the tilde symbol means the arithmetic average of the magnitude, Eq. (\ref{atilde}).

We applied Eq.~(\ref{conduciad}) to the classical test of simulating the propagation of a thermal wave in an homogeneous medium with $\rho=1$~g.cm$^{-3}$ and compared the results with those obtained using Eq.~(\ref{conducstd}). During the calculation, we keep the particles at rest so that the energy equation, Eq.~(\ref{energyqij}), reduces to the heat transport equation given by the expressions above. An initial point-like discontinuity in energy was seeded at the center of the box. To avoid numerical problems, the discontinuity was smoothed using a sharp Gaussian with characteristic width of a few times the smoothing length parameter. To achieve this we made use of the analytical expression for a spherically symmetric thermal-wave front in \cite{lan59}

\begin{equation}
u(r,t)=\frac{A}{(4\pi\alpha~t)^{\frac{3}{2}}}\exp\left(-\frac{r^2}{4\alpha~t}\right)+u_0\,,
\label{thermalwave}
\end{equation}

\noindent where $\alpha=\kappa/(c_v~\rho)$ is the thermal diffusivity and $c_v$ the specific heat capacity. The parameters in Eq.~(\ref{thermalwave}) were set to $A=10^5$~erg~cm$^3$~g$^{-1}$, $u_0= 10^3$~erg~g$^{-1}$,$c_v=9.1\cdot 10^6$~erg~g$^{-1}$~K$^{-1}$, and $\kappa= 3.9\cdot 10^9$~erg s$^{-1}$~cm$^{-1}$~K$^{-1}$. 

\begin{figure}
\includegraphics[angle=-90,width=\columnwidth]{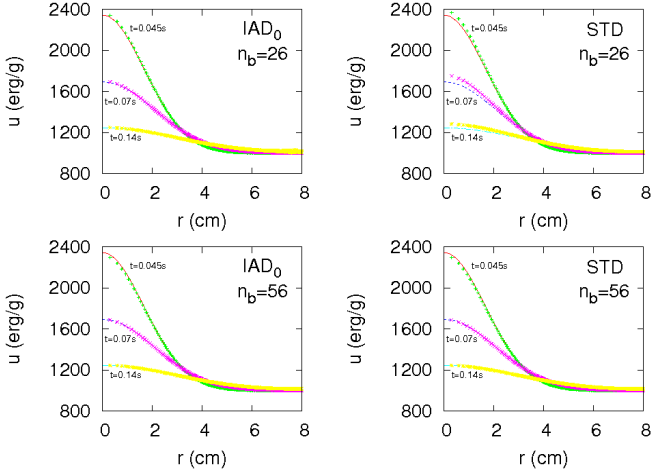}
\caption{Evolution of a thermal wave propagating through an ordered lattice of particles at different times. The profile of internal energy is shown for both SPH schemes and two different numbers of neighbors.}
\label{figure15}
\end{figure}

\begin{figure}
\includegraphics[angle=-90,width=\columnwidth]{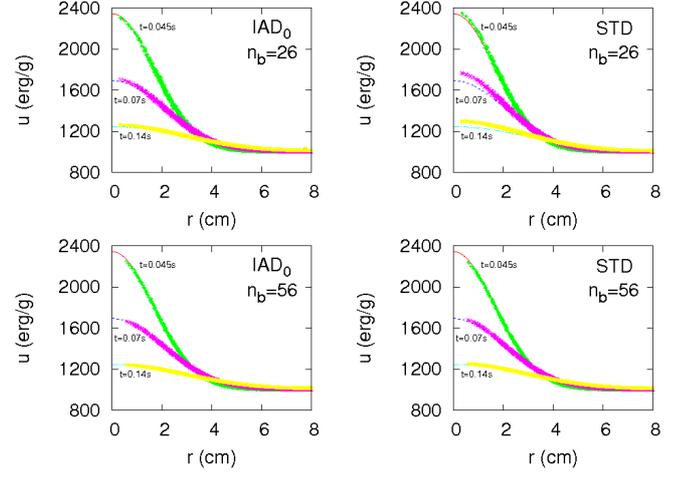}
\caption{Evolution of a thermal wave propagating through a pseudo-ordered distribution of particles at different times. The profile of internal energy is shown for both SPH schemes and two different numbers of neighbors.}
\label{figure16}
\end{figure}

To set the initial model, we put $47,707$ identical particles at the nodes of a 3D regular lattice. The initial energy profile was that given by Eq.~(\ref{thermalwave}) for $t=0.04$~s. The evolution of the thermal wave for different numbers of neighbors is depicted in Fig.~\ref{figure15}. As we can see, the tensor scheme is in closer agreement with the analytical solution when the number of neighbors is small, $n_b\simeq 25$. Nevertheless, above $n_b\simeq 50$ neighbors both schemes lead to similar results. To investigate the influence of the initial setting of particles we carried out a second calculation, this time using a pseudo-ordered lattice. The new particle distribution was obtained after randomly perturbing the 3D ordered lattice with a maximum perturbation amplitude $0.3\Delta$, where $\Delta$ is the lattice spacing. In this new calculation the mass of the particles was conveniently arranged to keep the density constant, at $\rho=1$~g~cm$^{-3}$, throughout the system. The maximum mass contrast between neighboring particles was lower than a factor of two. The results of this second calculation are summarized in Fig.~\ref{figure16}. As  we can see, the calculation with IAD$_0$ always reproduces more closely the evolution of the heat wave, especially for a small number of neighbors but also even when a moderate, $n_b\simeq 50$, amount of neighbors is chosen.          

\section{Computational issues}
\label{section5}
In this section we conducted a benchmark test to assess the impact of the calculation of the IAD$_0$ terms and the momentum and energy equations using Eqs.~(\ref{momentumLqij}-\ref{lagr8}). To do this we generated a 3D distribution of particles using a quasi-random Sobol distribution. Using the stretching technique, we then displaced the particles to follow a spherical density profile for the central core of a pre-collapse 15 M$_{\odot}$ star (see \cite{heger05}). The hydrocode is OMP-parallelized, and includes a Barnes-Hut octal-tree solver to calculating the gravitational force up to the quadrupolar term, and the Lattimer-Swesty EOS (\cite{latt91}). As we are only interested in the impact of the IAD$_0$ calculation on the overall wall-clock time, we set the velocity of the particles to zero at each time-step (frozen structure).

Table~\ref{table3} summarizes the outcome of the calculations. We varied the number of particles and the number of threads to gain insight into both the overhead and the scaling of the IAD$_0$ calculation. Table~\ref{table3} shows, from left to right, the number of particles, the number of threads, the tolerance parameter $\theta$, the wall-clock time per iteration for the standard calculation ($t^{TOT}_{STD}$), the wall-clock time spent by the code in the calculation of the standard momentum and energy equations ($t^{MOM}_{STD}$), the wall-clock time per iteration for the IAD$_0$ calculation ($t^{TOT}_{IAD_0}$), the wall-clock time spent by the code in the calculation of the IAD$_0$ coefficients ($t^{CAL}_{IAD_0}$), and the wall-clock time spent by the code in the calculation of the IAD$_0$ momentum and energy equations ($t^{MOM}_{IAD_0}$). The last column shows the IAD$_0$ overhead as a percentage, with respect to the standard simulation, and is calculated as

\begin{equation}
IAD_0 \text{ overhead} = \frac{t^{CAL}_{IAD_0}+t^{MOM}_{IAD_0}-t^{MOM}_{STD}}{t^{TOT}_{STD}}\times 100\,.
\end{equation}

\noindent All these times were averaged over 1000-2000 iterations for each calculation, which was always performed on the same 12-core AMD machine.

The parameter $\theta$ was used to control the criterion for choosing between opening a node in the tree (and using the particle-to-particle interaction to calculate gravity) and not opening the node (and using its global contribution via the multipolar expansion). The smaller the value of $\theta$, the slower the tree-walk, because more nodes are opened and the calculation gets closer to $N^2$ scaling. Typical values of $\theta$, those actually used in calculations, are between $0.5$ and $0.7$. Here, we selected two values ($0.3$ and $0.6$) to assess the relevance of the IAD$_0$-related calculations to the gravity evaluation, which is typically the most important bottleneck of the code.

\begin{figure}
\includegraphics[width=\columnwidth]{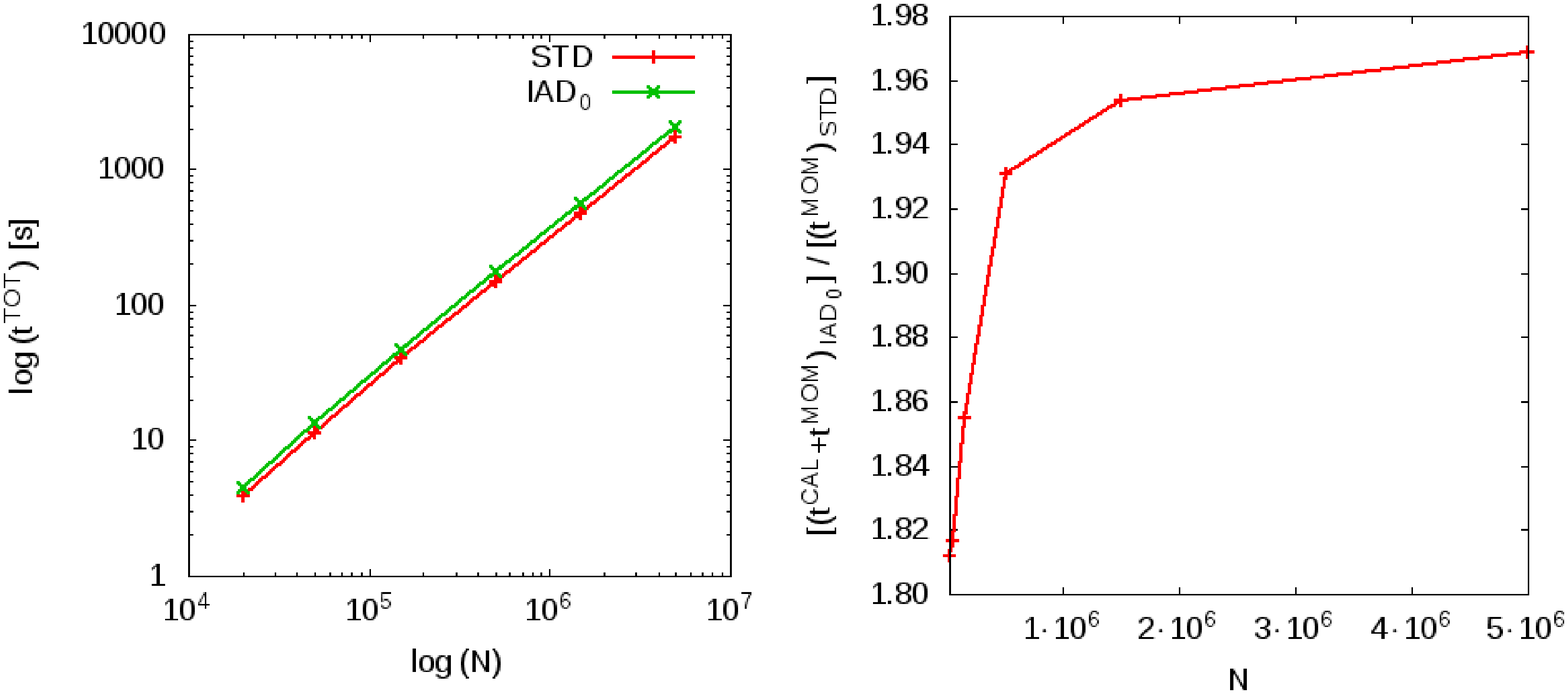}
\caption{Benchmarking of the impact of IAD$_0$ on the computational time. Left: Total wall-clock time per time-step in front of the number of particles. Right: Overload ratio of the IAD$_0$ related calculations (matrix coefficients + momentum and energy equations) to the standard momentum and energy calculation.}
\label{figure17}
\end{figure}

From these results, it can be seen that the inclusion of the IAD$_0$ scheme in single-threaded (serial) calculations contributes with an overhead of approximately $18\%$, independently of the number of particles used, which means that, as expected, it has a linear dependence on them. This can be seen in Fig.~\ref{figure17} (left), where it is clear that the inclusion of IAD$_0$ in the calculation, does not vary the slope of the curve. In the case of the standard calculation, the relationship between the wall-clock time and the number of particles is $log(t)\simeq log(N)^{1.0907}$, while for the IAD$_0$ calculation it is $log(t)\simeq log(N)^{1.0919}$. Even more, Fig.~\ref{figure17} (right) shows that the overhead ratio of the IAD$_0$-related calculations to their standard analogous tends to saturate at a factor slightly lower than 2. Noting that the momentum and energy calculations in the standard version of SPH takes around $20\%$ of the time, this result is consistent with the IAD$_0$ overhead. 

These are desirable properties as they mean that this section of the code will have at least the same behavior in terms of parallelization as the rest of the code, and that the overhead is limited independently of the number of particles. This can also be seen when we compare the single-threaded results with the multi-threaded ones. The wall-clock time for the IAD$_0$ sections in the 12-thread calculation is 9-10 times shorter than the time for the same sections in the serial work, which points to a good strong scaling and a considerable reduction in the IAD$_0$ overhead to $9\%$ of the overall wall-clock time.

The overhead reduction due to parallelization of course depends on how well the code is parallelized and the amount of it that remains serial. Nevertheless, the results that refer specifically to the IAD$_0$ sections show relevant information on the low impact that IAD$_0$ may have on the overall calculation when $these$ sections are parallelized.

Finally, it can also be seen from the last two rows of Table~\ref{table3} that the relevance of the IAD$_0$ sections is reduced to $\simeq 14\%$ when gravity becomes more dominant. This can be taken as a lower limit of the IAD$_0$ overhead in single-threaded calculations, as the parameter $\theta=0.3$ is rather extreme and is rarely used below that value.

\section{Conclusions}

We have checked the behavior of a novel SPH scheme where gradients are calculated using an integral approach. The main features of the technique, called IAD$_0$, were described in detail in paper I and summarized in Sec.~\ref{section2} of this paper. The main virtue of the approach relies in that the re-normalization of the derivatives appears naturally, without any degradation of the conservative properties that characterize the SPH technique. Another relevant feature is that the basic mathematical formalism of IAD$_0$ looks very similar to that of a standard SPH technique, making its implementation quite straightforward. As commented in paper I, matrix methods based on a similar, although not identical, formulation have been used in CDF for the past decade (see for instance \cite{dilts99}), but have never been previously applied to astrophysics.

Three test cases of considerable astrophysical interest were selected to validate IAD$_0$, as well as to detect its virtues and weaknesses. The performance of the method in describing the growth of the RT instability in a supernova remnant was analyzed in Sec.~\ref{remnant}, with the conclusion that IAD$_0$ provides a healthier development of the RT fingers. The method's success in handling hydrodynamical instabilities relies on the re-normalization imposed on the derivatives, thus confirming the results obtained in paper I using toy models. 

The second test was addressed especially to the applications of the method to describing stellar objects (approached as polytropes with different indexes and known analytical properties). One important question here relies in the treatment of the outer boundary of the object, which is often a controversial point in matrix methods \cite[]{oger07}. We have entirely explored this issue using several recipes to handle the surface of the polytropes. These include the use of tensor and vector formulations of IAD$_0$, as well as hybrid schemes that use the full matrix expressions in the interior but changes to {\sl vector-}IAD$_0$ near the surface, according to Eq.~(\ref{tauijselect}). The best results were obtained when the full tensor  IAD$_0$ scheme was used, especially for the most unstable polytrope with index n=5/2. Nevertheless, the amount of numerical noise stored as residual kinetic energy at equilibrium was larger for {\sl tensor-}IAD$_0$ than for STD.

The ability of the tensor method to handle the coalescence and further merging of stellar-like objects was checked in Sec.~\ref{merger}. Even though for this particular test the simulations did not show any clear advantage of the matrix method, it gave a good depiction of the coalescence process. Despite both schemes being conservative by construction, complete preservation of angular momentum is impeded by the use of the hierarchical cluster method in calculating gravity. Nevertheless, there are indications that the angular momentum-component orthogonal to the orbital plane behaves better in IAD$_0$ than in the standard scheme (Fig.~\ref{figure12}). The prompt product after the coalescence is a core surrounded by an extended diluted halo of particles moving at Keplerian velocity. The calculations show that the tensor method leads to a more homogeneous mixing of the material of both stars in the core region. As the numerical recipe for implementing AV in both calculations is exactly the same, we conclude that the different behavior is caused by the differences in the algorithm used to compute the kernel derivative. Therefore, it seems that IAD$_0$ is able to provide a more effective hydrodynamical mixing than the standard scheme, but still avoid the penetration of fluids in strong shocks, as suggested in paper I. These differences also affect the distribution of angular velocity in the core of the remnant, which in the STD calculation soon reaches rigid rotation but in the matrix one does not, for a similar elapsed time. A potential weakness of the tensor method in handling these kinds of configurations comes from the computation and further inversion of matrix $\mathcal T$ calculated with Eq. (\ref{tauijsph}). In the case of the isolation of a particle, as it could be for those belonging to the most diluted region of the domain, matrix $\mathcal T$ becomes singular leading to the complete breakdown of the simulation. In this sense, the matrix scheme is less robust than the standard one. Nevertheless, current SPH schemes usually include an algorithm to ensure that the number of neighbors of a given particle remain constant during the simulation. This algorithm is necessary to reliably compute the magnitude $\Omega$ in Eqs.~(\ref{momentumLqij}) and (\ref{energyqij}) and, when working properly, to avoid the singularity problems linked to matrix $\mathcal T$.

Another interesting features of the matrix method are its ability to avoid the pairing instability and the better handling of thermal conduction. The first one leads to less particle clustering, thus improving the quality of the interpolations; while in the second case, the results of Sec.~\ref{section4} suggest that diffusion-like equations can be handled by IAD$_0$ in a  better way than in the standard framework.

We also conducted a benchmark test to evaluate the impact of the inclusion of the IAD$_0$ formalism on the wall-clock time for a nominal calculation. According to Table~\ref{table3} and Fig.~\ref{figure17}, the computational overload introduced in a serial calculation by the re-normalization of the derivatives is low ($\simeq 20\%$), being practically independent of the total number of particles. The scaling of the code with the number of particles remains virtually untouched, and the IAD$_0$-related sections of the code show a good behavior in front of parallelization.

Therefore, the analysis of the astrophysical tests discussed in this paper support the main conclusions of paper I, where the basic implementation of the integral approach to the derivatives was discussed and checked. All this suggests that the use of the re-normalized, fully conservative IAD$_0$ approach to the SPH equations may improve, in general, the quality of the simulations in astrophysics with a little computational overload penalty.
            
The smaller amount of viscous dissipation shown by IAD$_0$, compared to STD for the same AV formalism, suggests that it could be of interest not only when handling hydrodynamic instabilities but also when simulating turbulence. Turbulence is at the heart of many astrophysical problems, being of especial relevance to understanding star formation (\cite{fed10}). In this respect, a full comparison of 3D astrophysical turbulence calculated with a variety of algorithms, both SPH and grid codes, was provided by \cite{kitsionas09}, \cite{pricefed10}, and \cite{kritsuk11}, with the result that both families of codes give similar results for an equivalent number of resolution elements in each direction in space. Nevertheless, these experiments also show that, owing to the higher dissipation, the scaling range of SPH codes is slightly shorter than that of grid-based codes, as demonstrated by \cite{kitsionas09}. Therefore, it may be of great interest to check the ability of the proposed IAD$_0$ scheme to handle astrophysical turbulence in the near future.

\section*{Acknowledgements}

The authors acknowledge useful conversations with Stephan Rosswog. This work has been funded by the Spanish MEC grants AYA2010-15685, AYA2011-23102, and the Swiss Platform for High-Performance and High-Productivity Computing (HP2C) within the {\em supernova} project. It was also supported by ESF EUROCORES Program Eurogenesis through the MICINN grant EUI2009-04167 and by DURSI of the Generalitat de Catalunya. The rendered SPH plots were made using the freely available $SPLASH$ code \citep{price07}.


\clearpage

\clearpage

\clearpage

\clearpage

\clearpage
\begin{table*}[h!]
\centering
\begin{tabular}{@{}llcccrrrrr@{}}
\hline
Model & Scheme & Particles       & Ejecta & Perturbation $\delta$ & $\vert~\Delta E\vert/E_0$ & $\vert\Delta r_{cm}\vert$/$\bar R$ & $\vert L\vert/(\sum_i\vert L_i\vert)$ \\
      &        & ($\times 10^6$) &        &  ($km.s^{-1}$)        &                           &                                    & \\ 
\hline
\hline
A$_1$&IAD$_0$         &1.46&rnd    &0  &$6.5~10^{-3}$ & $8~10^{-5}$   & $10^{-4}$ \\
A$_2$&STD             &1.46&rnd    &0  &$1.3~10^{-2}$ & $8.2~10^{-5}$ & $4~10^{-4}$ \\
B$_1$&IAD$_0$         &1.46&rnd    &500&$6.5~10^{-3}$ & $7.5~10^{-5}$ & $5.5~10^{-5}$ \\
B$_2$&STD             &1.46&rnd    &500&$1.3~10^{-2}$ & $8.4~10^{-5}$ & $4~10^{-4}$ \\
B$_3$&IAD$_0$ (vector)&1.46&rnd    &500&$7~10^{-3}$   & $7.5~10^{-5}$ & $3~10^{-4}$ \\
C$_1$&IAD$_0$         &1.46&Stretch&0  &$6.5~10^{-3}$ & $9~10^{-12}$  & $4~10^{-12}$ \\
C$_2$&STD             &1.46&Stretch&0  &$1.3~10^{-2}$ & $7~10^{-11}$  & $9~10^{-9}$ \\
\hline
\end{tabular}
\caption{Comparison between several magnitudes at $t=698$~yr of evolution of the SNR computed using IAD$_0$ and STD schemes. The deviation of the center of mass is normalized to $\bar R=2.5$~pc. $\sum_i\vert L_i\vert=\sum_i\left(\vert Lx_i\vert+\vert Ly_i\vert+\vert Lz_i\vert\right)$.}
\label{table1}
\end{table*}

\begin{table*}[h!]
\centering
\begin{tabular}{@{}llrrrrrrrr@{}}
\hline
Model & scheme& Mass (M$_{\sun})$&index n& $\rho_c$ (g.cm$^{-3})$& $R_s$~(cm)&$\Delta\rho_c/\rho_c$& $\Delta R_s/R_s$ & $E_{kin}/E_{tot}$&$\vert~\Delta E\vert/E_0$\\ 
\hline
\hline
E$_1$&IAD$_0$        &0.6 &3/2&$3.3~10^6$&$8~10^8$&$1~10^{-3}$&0.02 &$6~10^{-4}$ & $2~10^{-3}$  \\
E$_2$&STD            &0.6 &3/2&$3.3~10^6$&$8~10^8$&$6~10^{-2}$&0.11 &$3~10^{-5}$ & $4~10^{-4}$   \\
E$_3$&IAD$_0$(vector)&0.6 &3/2&$3.3~10^6$&$8~10^8$&$3~10^{-2}$&0.14 &$4~10^{-5}$ & $2~10^{-4}$  \\
F$_1$&IAD$_0$        &1.35&5/2&$1.9~10^9$&$2~10^8$&0.33       &0.04 &$8~10^{-4}$ & $2~10^{-3}$  \\
F$_2$&STD            &1.35&5/2&$1.9~10^9$&$2~10^8$&0.42       &0.11 &$2~10^{-4}$ & $2~10^{-3}$   \\
F$_3$&IAD$_0$(vector)&1.35&5/2&$1.9~10^9$&$2~10^8$&$5~10^{-3}$&0.24 &$5~10^{-5}$ & $3~10^{-4}$  \\
\hline
\end{tabular}
\caption{Main features of polytropes with polytropic indexes $n=3/2$ and $n=5/2$, respectively. The analytical value of the central density $\rho_c$ and surface radius $R_s$ are given in columns 4 and 5. The relative errors between the numerical and analytical estimations are provided in columns 7 and 8. The kinetic energy stored as numerical noise (normalized to the internal energy) and the level of energy conservation at the last calculated model are given in columns 9 and 10.}
\label{table2}
\end{table*}

\begin{table*}[h!]
\centering
\begin{tabular}{@{}lccrrrrrr@{}}
\hline
Particles & Threads & $\theta$ & $t^{TOT}_{STD}$ & $t^{MOM}_{STD}$ & $t^{TOT}_{IAD_0}$ & $t^{CAL}_{IAD_0}$ & $t^{MOM}_{IAD_0}$ & IAD$_0$ overhead\\
\hline
\hline
20,000    & 1  & 0.6 & 3.84    & 0.85   & 4.51    & 0.55   & 0.99   & 17.97\% \\
50,000    & 1  & 0.6 & 11.27   & 2.45   & 13.32   & 1.40   & 3.05   & 17.75\% \\
150,000   & 1  & 0.6 & 40.29   & 8.12   & 47.18   & 4.80   & 10.26  & 17.22\% \\
500,000   & 1  & 0.6 & 148.87  & 28.24  & 175.40  & 18.11  & 36.42  & 17.66\% \\
1,500,000 & 1  & 0.6 & 474.68  & 88.50  & 558.69  & 56.33  & 116.56 & 17.78\% \\
5,000,000 & 1  & 0.6 & 1750.95 & 341.29 & 2082.26 & 198.11 & 473.78 & 18.88\% \\
\hline
500,000   & 12 & 0.6 & 27.27   & 3.73   & 30.37   & 1.73   & 4.33   & 8.50\%  \\
5,000,000 & 12 & 0.6 & 311.51  & 38.08  & 347.76  & 18.23  & 48.70  & 9.26\%  \\
\hline
50,000    & 1  & 0.3 & 14.28   & 2.47   & 16.46   & 1.42   & 3.09   & 14.29\% \\
150,000   & 1  & 0.3 & 48.25   & 8.16   & 55.10   & 4.81   & 10.22  & 14.24\% \\ 
\hline
\end{tabular}
\caption{Average wall-clock time for the different benchmark tests and sections of the code relevant to the IAD$_0$ calculation. $\theta$ is the parameter for opening the nodes in the tree when calculating gravity. $TOT$, $MOM$, and $CAL$ stand for total time per iteration, momentum and energy equation calculations, and IAD$_0$ terms calculation, respectively. All times are expressed in seconds.}
\label{table3}
\end{table*}

\end{document}